\documentclass[a4paper]{article}

\usepackage{graphicx}
\usepackage{mathrsfs}  
\usepackage{float}

\begin{document}

\title{Empirical large--scale extension of Yakhot's model of strong turbulence}
\author{Ch.  Renner} 
\date{July 16, 2024} 
\maketitle

\begin{abstract}
We propose an empirical extension of Yakhot's model of strong turbulence \cite{Yakhot} that correctly describes the statistics of longitudinal velocity increments not only in the inertial range but also for larger scales up to the system length scale $L$. Two additional parameters are introduced to the original equation for the probability density function of velocity increments. These parameters can be motivated by physical arguments and their values are fixed by large scale boundary conditions. The resultant model ensures correct convergence of structure functions at the system length scale, including vanishing slopes at $L$, and shows good agreement with experimental data.
\end{abstract}

\section{Introduction}

Even though the first theory of fully developed turbulence dates back to 1941 \cite{K41}, the statistical properties of velocity fluctuations in turbulent fluid motion remain a challenging open problem. Velocity fluctuations on a certain length scale $l$ are usually investigated by means of the longitudinal velocity increment $v(l)$, the difference of the velocities at two points in space separated by the distance $l$: $v(l) = w(x+l)-w(x)$ (here, $w$ is the component of the velocity field in direction of the separation vector $\bf l$). The statistics of $v$ is often characterized by means of the moments $\mathscr{S}_n(l) \; = \; <v(l)^n>$, the so-called structure functions.

A commonly accepted model for the structure functions has not yet been found. The only exception is the third order function for which an exact relation can be derived directly from the Navier–Stokes equation, Kolmogorov’s famous four–fifths law. Neglecting the dissipative term, the third order structure function is a linear function of the scale $l$
\begin{equation}
\mathscr{S}_3(l) \; = \; - \, \frac{4}{5} \, \mathcal{E} \, l.
\label{fourFifthsReduced}
\end{equation}
where $\mathcal{E}$ is the mean rate of energy dissipation within the flow. (\ref{fourFifthsReduced}) is valid for scales $l$ much smaller than the system size $L$ and larger than the dissipation length scale $\eta$ of the flow configuration, the so--called inertial range of length scales. 

Inspired by this result models of fully developed turbulence are usually based on the hypothesis that in the inertial range structure functions follow power laws in $l$
\begin{equation}
\mathscr{S}_n(l) \; \propto \; l^{\zeta_n}, 
\label{scaling}
\end{equation}
where the $\zeta_n$ are the so--called scaling exponents. Several models for these exponents have been developed in recent decades, from the simple linear relation $\zeta_n = \frac{n}{3}$ proposed by Kolmogorov in 1941 \cite{K41} to, amongst others, functions of second \cite{K62} and third \cite{Lvov} order in $n$. 

In 1998, V. Yakhot proposed a model of fully developed turbulence in the limit of very large Reynolds number which is based on an analytical treatment of the Navier–Stokes equation \cite{Yakhot}. A result of this model is the prediction of a closed--form expression for the scaling exponents $\zeta_n$ which can be shown to be a generalization of several of the most important models of turbulence \cite{Mine}. 

Remarkably enough, this prediction for the scaling exponents replicates an earlier results by B. Castaing \cite{Castaing} who showed that very general considerations allow to identify a quantity conserved along scales which can be thought of as the analogon of temperature in thermodynamics. From this, he was able to derive a closed--form expression for the scaling exponents which preempts Yakhot's.

Unlike many others, Yakhot's model is not restricted to the inertial range but also correctly describes important aspects of the transition to large length scales. In particular, the model correctly captures the decline\footnote{Strictly speaking this statement refers to absolute values: Odd--order structure functions are negative in the inertial range, cf. eq. (\ref{fourFifthsReduced}), and so actually {\it increase} for larger scales to finally reach zero from below.\label{signOfS3}} of odd--order structure functions to zero at the system length scale. Yet, the model misses out on a few other aspects of the transition to large scales. One of the aspects not described correctly is the convergence to constant levels of {\it even--order} structure functions at the system scale $L$. Also, structure functions approach the system scale $L$ with non--zero slopes in Yakhot's model which runs contrary to physical intuition and experimental data.

In this paper we propose a straightforward empirical extension of Yakhot's model which also captures these features. Two additional parameters are introduced into the original partial differential equation for the evolution of the velocity increment's probability density function. These parameters can be motivated by physical considerations and their values are fixed by large scale boundary conditions. The resultant structure functions show good agreement with experimental data. The proposed extension is empirical, but we believe that the rationale underlying the model and the results obtained with it are sound, making it a prospective starting point for further experimental and theoretical investigations.

\section{Large Scale Boundary Conditions}\label{boundaries}

The large scale boundary conditions for the structure functions are not dependent on a specific model but can be derived from rather general considerations and assumptions. We start by defining the system length scale $L$ as the scale at which the velocity fluctuations $w$ have de--correlated:
\begin{equation}
	\left< \, w(x+l) \, w(x) \, \right> \; = \; 0. \qquad \; \textrm{for } l \,  \geq \, L \label{Lcondition}
\end{equation}
It should be noted that the system scale defined by (\ref{Lcondition}) is not identical with the often used integral length scale, which is also usually denoted by $L$ (see \cite{ift} and references therein for the definition of the integral scale). The integral scale marks the upper end of the scaling range and is significantly smaller than the system scale as used in this paper.

For the second order structure function, the definition (\ref{Lcondition}) implies that at length scale $L$ (and larger):
\begin{eqnarray}
	\mathscr{S}_2(L) \; =  \; \left< \left( \, w(x+L)-w(x) \, \right)^2 \right> 
	\;  =  \; 2 \, \sigma^2,  \label{s2LargeScaleValue}
\end{eqnarray}
where $\sigma = \sqrt{\left< w^2 \right>}$ is the rms of the velocity fluctuations $w$.

We furthermore assume that velocity fluctuations $w$ on large scales follow normal distributions. Thus, the increment $v(L)$ is the difference of two uncorrelated normal processes and hence a normal random processes itself with a variance given by (\ref{s2LargeScaleValue}). The large scale boundary levels of structure functions follow from general properties of the normal distribution as:
\begin{eqnarray}
	\mathscr{S}_n(L) \; = \; \left< v(L)^n \right>  \; = \;
	  \left\{
    \begin{array}{ll}
      (n-1)!! \; (2\sigma^2)^\frac{n}{2} \qquad & \textrm{if $n$ even,} \\ \\
      0   & \textrm{if $n$ odd}
    \end{array}
  \right.
	\label{largeScaleValues}
\end{eqnarray}
where $(n-1)!! = 1 \cdot 3 \cdot \ldots \cdot (n-3) \cdot (n-1)$.

The second set of boundary conditions, expected both from physical intuition as well as experimental data, is vanishing first order derivatives at the system scale $L$ for structure functions of all orders:
\begin{equation}
	\left. \frac{\partial}{\partial l} \, \mathscr{S}_n(l) \, \right|_{l=L}\; = \; 0 \qquad \forall \; n. \label{zeroSlopeCondition}
\end{equation}

\section{Yakhot's Model of Strong Turbulence}\label{yakhotsModel}

Following ideas by A.M. Polyakov \cite{Polyakov}, Yakhot proposed a model of fully developed turbulence in the limit of very large Reynolds number which is based on an analytical treatment of the Navier--Stokes equation \cite{Yakhot}. For the case of turbulence driven by Gaussian velocity fluctuations on large scales he derived a partial differential equation for the evolution of the probability density function $p(v,l)$ of the velocity increment $v$ in scale $l$:

\begin{equation}
	B \frac{\partial p}{\partial l} \, - \, \frac{\partial}{\partial v} 
	\left\{ v \frac{\partial p}{\partial l} \right\} \; = \; - \frac{A}{l} 
	\frac{\partial}{\partial v} \big\{ \, v \, p \, \big\} \, + \, \frac{\sigma }{L} 
	\frac{\partial^2}{\partial v^2} \big\{ \, v \, p \, \big\}, \label{originalYakhot}
\end{equation}
where $A$ and $B$ are two parameters that are not fixed by the theory. A remarkable feature of this equation, amongst others \cite{Mine}, is the occurrence of the system scale $L$ in the second term on the right--hand side, i.e. the fact that large--scale effects are explicitly taken into account. It is this term that describes the decline\footref{signOfS3} of odd-order structure functions to zero at $L$.

For what follows we will discuss this equation in the dimensionless variables $r$ and $u$ defined as:
\begin{equation}
	r = l/L, \qquad  u = v/\sigma.
\label{uAndR}
\end{equation}
With these definitions the equation for the probability density function is:
\begin{equation}
	B \frac{\partial p}{\partial r} \, - \, \frac{\partial}{\partial u} 
	\left\{ u \frac{\partial p}{\partial r} \right\} \; = \; - \frac{A}{r} 
	\frac{\partial}{\partial u} \big\{ \, u p \, \big\} \, + \, 
	\frac{\partial^2}{\partial u^2} \big\{ \, u p \, \big\}. \label{PDFEquation}
\end{equation}

By multiplication of eq. (\ref{PDFEquation}) with $u^n$ and subsequent integration with respect to $u$, the equation for the dimensionless structure functions $S_n(r) = \mathscr{S}_n(r)/\sigma^n = \int u^n p(u) du$ can be derived:
\begin{eqnarray}
	r \frac{\partial}{ \partial r} S_{n}(r) \;  & =  & \; \zeta_{n} \, S_{n}(r) \, +  \,z_{n} \, r \, S_{n-1}(r), \label{SunEquation} \\
	\zeta_{n} \;  & =  & \; \frac{An}{B+n} \label{zetaN1},  \\
	z_{n} \;  & =  & \; \frac{n(n-1)}{B+n}. \label{zN1}
\end{eqnarray}
In the limit of small scales $r  \ll 1$, the second term on the rhs of equation (\ref{SunEquation}) becomes negligible. In this limit, the solutions of the resulting equation are simple power laws:
\begin{equation}
	S_{n}(r) \; \propto \; r^{\zeta_{n}}. \label{SmallScaleSuns}
\end{equation}
The four--fifths law (\ref{fourFifthsReduced}) imposes the condition $\zeta_{3}=1$ on the scaling exponents which, inserted into equation (\ref{zetaN1}), leads to:
\begin{equation}
	A \; = \; \frac{B+3}{3}. \label{AvonB}
\end{equation}
The $\zeta_n$ and $z_n$ can thus be expressed as a function of only one yet unknown parameter, either $B$ as in Yakhot's original work \cite{Yakhot}, or $A$ as for the most part of this paper:
\begin{eqnarray}
	\zeta_{n} \; & = & \; \frac{n}{3} \frac{B+3}{B+n} \; = \; \frac{A \, n}{3 \, (A-1) \, + \, n}. \nonumber \\
	z_n \; & = & \frac{n \, (n-1) }{B \, + \, n} \; = \; \frac{n \, (n-1) }{3 \, (A-1) \, + \, n}.  \label{ZetaN}
\end{eqnarray}
The choice of $A$ as independent parameter is, amongst other things, motivated by the fact that it can be interpreted as the limit of $\zeta_n$ for $n\rightarrow \infty$:
\begin{eqnarray}
	\lim_{n\rightarrow \infty} \, \zeta_{n} \; = \; \lim_{n\rightarrow \infty} \, \frac{A}{\frac{3 \, (A-1)}{n} \, + 1} \; = \; A.  \label{aAsZetaLimit}
\end{eqnarray}
It is worthwhile noting again that the scaling exponents in Yakhot's model are in line with an earlier result by B. Castaing \cite{Castaing}.

For the second order structure function the last term on the rhs of eq. (\ref{SunEquation}) vanishes on all scales (as $S_{1}(r)=0$) and the general solution for $S_2$ is the power law (\ref{SmallScaleSuns}). The constant of integration can be determined from the large scale boundary condition (\ref{largeScaleValues}) yielding:
\begin{equation} 
	S_2(r) \; = \; 2 \, r^{\zeta_2}. \label{YakhotsSecondOrder}
\end{equation}
Inserting this result into the equation for the third order structure function leads to:
\begin{eqnarray}
	r \frac{\partial}{\partial r}S_{3}(r) \; & = & \; \zeta_{3} \, S_{3}(r) + z_{3} \, r \, S_{2}(r) \nonumber \\
	  & = & \; S_{3}(r) \, + \, 2 \, z_{3} \, r^{\zeta_{2}+1} . 
\end{eqnarray}
The general solution for this equation is:
\begin{eqnarray}
	S_{3}(r) \; & = & \; K_{3} \, r + 2 \, \frac{z_{3}}{\zeta_{2}} \, r^{\zeta_{2}+1}.
\end{eqnarray}
The integration constant $K_{3}$ can be determined from the four--fifths law (\ref{fourFifthsReduced}) which in the dimensionless variables $u$ and $r$ can be written as
\begin{equation}
	{S}_3 \left( r \ll 1 \right) \; = \; 
	- \, \frac{4}{5} \, \epsilon \, r, \label{fourFifthWoDimensions}
\end{equation}
where $\epsilon = \frac{\mathcal{E} L}{\sigma^3}$ is the dimensionless mean rate of energy dissipation. The final result for the third order structure function is:
\begin{eqnarray}
	S_{3}(r) \; & = & \; - \frac{4}{5} \, \epsilon \, r \, + \, 2 \, \frac{z_{3}}{\zeta_{2}} \, r^{1+\zeta_{2}} \nonumber \\
	& = & \; - \frac{4}{5} \, \epsilon \, r \, + \, 18 \, \frac{B+2}{\left( B+3 \right)^2} \, r^{1+\zeta_{2}} . \label{Su3}
\end{eqnarray}
The parameter $B$ can be determined from the condition that the third order moment vanish at the integral length scale, i.e. 
$S_{3}(r=1)=0$:
\begin{equation}
	\frac{4}{5} \,\epsilon \; = \; 18 \, \frac{B+2}{\left( B+3 \right)^2} 
	\label{BBestimmung}
\end{equation}
From dimensional arguments Yakhot inferred that $\epsilon \approx 1$. Equation (\ref{BBestimmung}) is then solved by $B \approx 18.5$ which corresponds to $A \approx 7.2$. For these values the scaling exponents (\ref{ZetaN}) are, within the errors, indistinguishable from experimental values \cite{Yakhot}.

\section{Shortcomings of Yakhot's Model}\label{shortcomings}
Yakhot's model stands out in several aspects: The Castaing--Yakhot scaling exponents (\ref{ZetaN}) can be shown to generalize several of the most relevant theories of turbulence \cite{Mine} and unlike many other models it captures the decline of odd--order structure functions to zero. Yet, several other features of structure functions are not described correctly. These aspects will in what follows be discussed by benchmarking the predictions of Yakhot's model against experimental data. These were measured in a cryogenic axisymmetric helium gas jet at a Reynolds number of approx. $4 \cdot 10^5$. Details on on the experimental setup and the data can be found in \cite{data, transition} and appendix \ref{app}.

\begin{figure}[ht]
	\centering
		\includegraphics[width=1\textwidth]{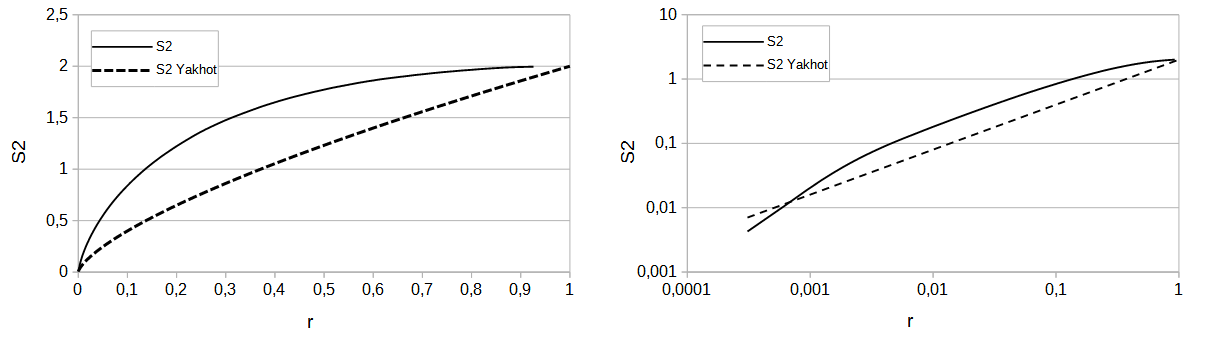}
	\caption{Empirical secon--order structure function (solid line, see appendix \ref{app} for details) in comparison to Yakhot's model (dashed line) in linear (left) and logarithmic (right) scales.}
	\label{yakhotsS2}
\end{figure}

A comparison of the model's prediction (\ref{YakhotsSecondOrder}) for the second order structure function with experimental data is shown in figure \ref{yakhotsS2}. When parameterized to match the large scale boundary condition $S_2(r=1)=2$, the model fails to describe the function at smaller scales. 

\begin{figure}[ht]
	\centering
		\includegraphics[width=1\textwidth]{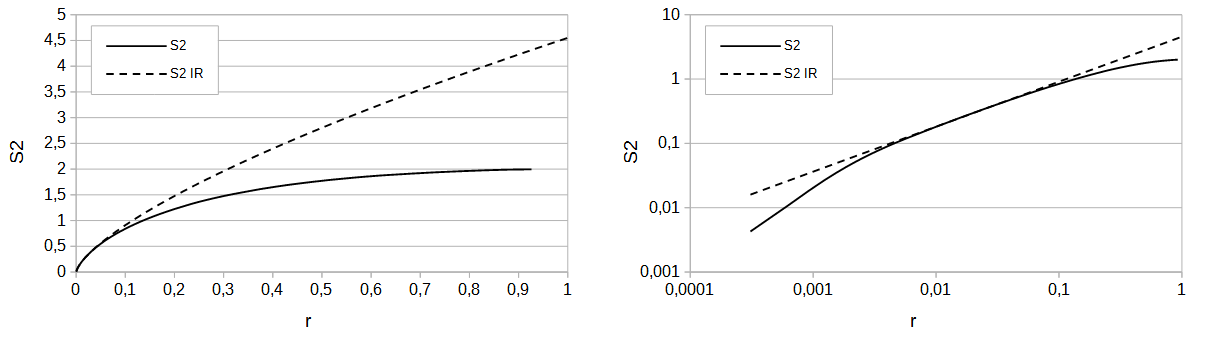}
	\caption{Empirical second--order structure function (solid line) in comparison to Yakhot's model with adjusted constant of integration (dashed line) in linear (left) and logarithmic (right) scales.}
	\label{yakhotsS2scaled}
\end{figure}

A better fit to the inertial range can be obtained by adjusting the constant of integration as shown in figure (\ref{yakhotsS2scaled}). However, the necessary increase of the constant of integration (from $2$ as in eq. (\ref{YakhotsSecondOrder}) to approximately $4.5$) leads to a massive overestimation of the large scale level $S_2(r=1)$. It is clearly not possible to correctly describe both the scaling in the inertial range as well as the large scale boundary level of the second order structure function.

\begin{figure}[ht]
	\centering
		\includegraphics[width=1\textwidth]{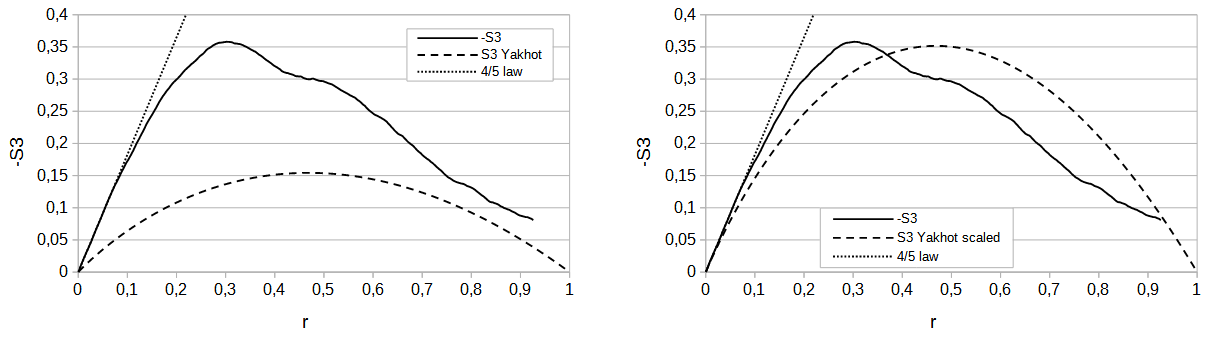}
	\caption{Empirical (negative) third--order structure function (solid line) in comparison to Yakhot's model (dashed line) in original (left) and adjusted (right) parameterization. Dotted lines indicate linear fits according to the four-fifths law (\ref{fourFifthsReduced}).}
	\label{yakhotsS3}
\end{figure}

Results for the third order structure functions also exhibit significant deficiencies. Figure \ref{yakhotsS3} displays the negative of $S_3$ in comparison to the predictions of Yakhot's model. In original parameterization (left) according to eq. (\ref{fourFifthWoDimensions}) the model, while correctly declining to zero at $r=1$, fails to match the function at all other scales, including the inertial range where $S_3$ should exhibit a linear dependence on scale according to the four--fifths law. A closer investigation reveals that this is due to the fact that the approximation $\epsilon \approx 1$ for the dimensionless rate of energy dissipation as used for the derivation of eq. (\ref{fourFifthWoDimensions}) does not hold. From fits to experimental data as indicated by the dotted lines in figure \ref{yakhotsS3} we in fact obtain a value of $\epsilon \approx 2.3$. Using this value and the scaled constant of integration for $S_2$, the parameterization of Yakhot's model can be adjusted and is then found to be in better agreement with the experimental data (right).

But also with the adjusted parameterization the model misses out on an important aspect, the linear dependence on scale in the inertial range according to the four--fifths law. In comparison to a linear fit according to that law (figure \ref{yakhotsS3}, dotted lines) it becomes evident that the influence of the second order structure function, which causes the deviation from the linear scaling and the eventual decline towards zero, sets in too early, i.e. at too small length scales. Considering that the four--fifth law is the only exact relation known this constitutes a serious drawback of the model.

Moreover, for any parameterization the structure functions will in this model always exhibit finite first order derivatives at the system length scale $L$. This is in contradiction to experimental data as can be seen from the results presented here for the second order structure function (while results for the third order function are less clear owing to higher statistical noise, see appendix \ref{app}).

\section{Extended Model}\label{ExtendedModel}
\subsection*{Specification}
We seek to address the shortcomings of Yakhot's model through an extension of the original equation (\ref{PDFEquation}) by as simple as possible elementary functions. As we are interested in the large scale behaviour of the model it seems natural to concentrate on the second term on the right--hand side of eq. (\ref{PDFEquation}), the term describing the influence of large scales.

Experimental results based on the theory of Markov processes \cite{howItAllStartet, myMarkovPaper, ift} motivate a first extension. In these analyses, the evolution in scale of the longitudinal velocity increment was described by means of the Fokker--Planck--equation, a partial differential equation for the pdf $p(u,r)$. In this framework, the stochastic component of the process is described by the so--called diffusion coefficient $D_2(u,r)$ which can be a function of both $r$ and $u$. Analyzing experimental data, the diffusion coefficient was found to have constant, linear and second order terms in $u$.

It could be shown \cite{DavoudiTabar} that equation (\ref{originalYakhot}) can approximately be converted to an equivalent Fokker--Planck--equation. The resultant effective diffusion coefficient comprises linear and second order terms in $u$, but lacks a constant term. This is not only inconsistent with experimental results but also counter--intuitive from a theoretical point of view: Diffusion coefficients that contain only terms of order $u$ or higher would imply that the stochastic variable $u$ will never change its sign\footnote{Ignoring the effect of the drift term.}. The corresponding pdf $p(u,r)$ would accordingly be restricted to one half of the real axis\footnote{The best known example for such a process is geometric Brownian motion for which the diffusion coefficient is of second order and the corresponding distribution is lognormal.} which is clearly not a reasonable assumption for the probability density function of velocity increments. Further compelling evidence for the existence of such an additive noise term in the Fokker--Planck equation was recently given in a study based on the integral fluctuation theorem \cite{ift}.

While the pdf derived by Yakhot, equation (\ref{originalYakhot}), is not a Fokker--Planck--equation we deem it reasonable based on above considerations to introduce an effective constant (in $u$) diffusion term. In the Fokker-Planck--equation such a term takes on the form $D(r) \, \frac{\partial^2}{\partial u^2} p(u,r)$ which nicely integrates into the large scale term in (\ref{PDFEquation}). For the purpose of modeling the large scale behaviour, we find that $D$ can be assumed constant\footnote{A more detailed investigation shows that this assumption does not hold for small scales.\label{dIsNotConstantFootnote}} in $r$.

A further adjustment can be inferred from the observations discussed in section \ref{shortcomings}, in particular the fact that the third order structure function deviates from the four-fifth scaling "too early", i.e. at too small length scales. In order to address this we propose to modify the large scale term by multiplying it with a function $c(r)$ that goes to zero for small scales:
\begin{equation}
	\lim_{r \rightarrow 0} \, c(r) \; = \; 0. \label{cGoesToZero}
\end{equation}

Based on these considerations we propose the following extension to Yakhot's model:
\begin{equation}
	B \frac{\partial p}{\partial r} \, - \, \frac{\partial}{\partial u} 
	\left\{ u \frac{\partial p}{\partial r} \right\} \; = \; - \frac{A}{r} 
	\frac{\partial}{\partial u} \Big\{ u p \Big\} \, + \, 
	\frac{\partial^2}{\partial u^2} \Big\{ \, \big( \, u \, c(r) + D \, \big) \, p  \, \Big\}. \label{pdfEquationExtended}
\end{equation}
By multiplication with $u^n$ and integration with respect to $u$ the equation for $S_n(r)$ is obtained:
\begin{eqnarray}
	r \frac{\partial}{ \partial r} S_{n}(r) \;  =  \; \zeta_{n} \, S_{n}(r) \, +  \,z_{n} \, r \, \Big\{ \, c(r) \, S_{n-1}(r) \, + \, D \, S_{n-2}(r) \, \Big\}, \label{SunEquationExtended} 
\end{eqnarray}
with $\zeta_n$ and $z_n$ as in (\ref{ZetaN}).

\subsection*{Exploiting the Large Scale Boundary Conditions}
The additional degrees of freedom can be used to enforce the condition of vanishing first order derivatives of the structure functions at the system length scale $r=1$. For odd--order structure functions the condition is.
\begin{eqnarray}
	0 \; & \stackrel{!}{=} & \; \left. r \frac{\partial}{ \partial r} S_{2n+1}(r) \right|_{r=1} \nonumber \\[0.1cm]
	& = & \; \zeta_{2n+1} \, S_{2n+1}(1) \, +  \,z_{2n+1} \, 1 \, \Big\{ \, c(1) \, S_{2n}(1) \, + \, D \, S_{2n-1}(1) \, \Big\} \nonumber \\[0.1cm]
	& = & \; \,z_{2n+1} \, c(1) \, S_{2n}(1). \label{cLargeScaleCondition}
\end{eqnarray}
The last line follows from the fact that the odd--order functions $S_{2n+1}$ and $S_{2n-1}$ are zero at the system scale, see eq. (\ref{largeScaleValues}).

In order for the first order derivative at the system scale to vanish also the remaining term involving $S_{2n}$ in (\ref{cLargeScaleCondition}) needs to be zero. This implies that:
\begin{equation}
	c(r=1) \; = \; 0. \label{cAtOne}
\end{equation}
The simplest elementary function fulfilling both condition (\ref{cAtOne}) as well as (\ref{cGoesToZero}) is:
\begin{equation}
	c(r) \; = \; r \, (1-r) \, C, \label{cVonR}
\end{equation}
with constant parameter $C$.

The parameter $D$ is fixed by imposing the condition of vanishing first order derivatives on even--order structure functions:
\begin{eqnarray}
	0 \; & \stackrel{!}{=} & \; \left. r \frac{\partial}{ \partial r} S_{2n}(r) \right|_{r=1} \nonumber \\
	& = & \; \zeta_{2n} \, S_{2n}(1) \, +  \,z_{2n} \, 1 \, \Big\{ \, c(1) \, S_{2n-1}(1) \, + \, D \, S_{2n-2}(1) \, \Big\} \nonumber \\[0.1cm]
	& = & \; \frac{2An}{B+2n} \, S_{2n}(1) \, + \, \frac{2n(2n-1)}{B+2n} \, D \, S_{2n-2}(1)  \nonumber \\[0.1cm]
	& = & \; \frac{2n}{B+2n} \, S_{2n-2}(1) \,  \left\{ \, A \, \frac{S_2n(1)}{S_{2n-2}(1)}  \, + \, D \, (2n-1) \, \right\} \nonumber \\[0.1cm]
	& = & \; \frac{2n}{B+2n} \, S_{2n-2}(1) \, (2n-1) \, \Big\{ \, 2 \, A  \, + \, D \, \Big\}, \nonumber 
\end{eqnarray}
where in the last step we used the equality $\frac{S_2n(1)}{S_{2n-2}(1)} = 2 (2n-1)$ which follows from the large scale boundary condition (\ref{largeScaleValues}). We obtain:
\begin{equation}
	D \; = \; - 2 \, A. \label{valueOfD}
\end{equation}
Using this relation to replace $D$ in eq. (\ref{SunEquationExtended}) we finally obtain:
\begin{eqnarray}
	\frac{\partial}{ \partial r} S_{n}(r) \; & = & \; \frac{\zeta_{n}}{r} \, S_{n}(r) \, + \, z_{n} \, c(r) \, S_{n-1}(r) \, - \, 2(n-1) \,\zeta_n \, S_{n-2}(r). 
	\label{extendedSunWithoutD}
\end{eqnarray}

\subsection*{Solutions and Fit to Experimental Data}
\begin{figure}[H]
	\centering
		\includegraphics[width=1\textwidth]{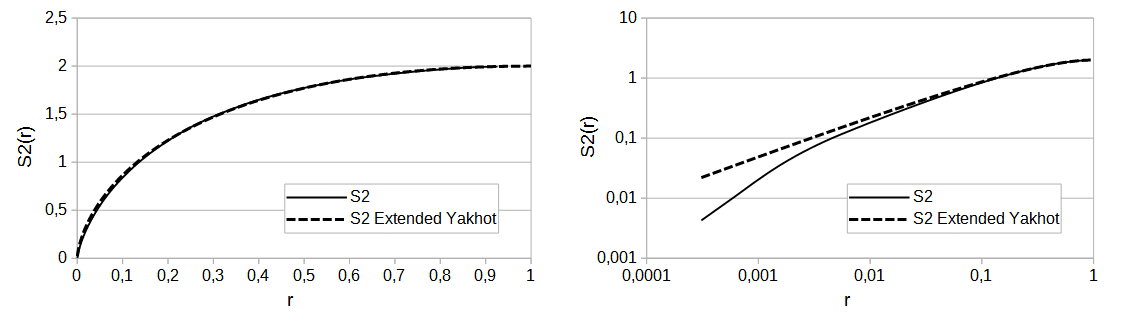}
	\caption{Second--order structure function as given by eq. (\ref{extS2Solution}) for $\zeta_2=0.7$ (dashed lines) in comparison to experimental data (solid lines) in linear (left) and logarithmic (right) scale.}
	\label{su2Fit}
\end{figure}

For order $n=2$, eq. (\ref{extendedSunWithoutD}) simplifies to:
\begin{eqnarray}
	r \frac{\partial}{ \partial r} S_{2}(r) \;  
	& = &  \; \zeta_{2} \, S_{2}(r) \, - \, 2 \, \zeta_{2} \, r.	\label{extS2Eq} 
\end{eqnarray}
(Note that $S_0(r)=1$.) With boundary condition $S_2(1)=2$ this is solved by:
\begin{eqnarray}
	S_{2}(r) \;  & = &  \; \frac{2}{1-\zeta_2} \, \left( \, r^{\zeta_2} \, - \, \zeta_2 \, r \, \right).  \label{extS2Solution} 
\end{eqnarray}
A comparison with experimental data shows good agreement over a wide range of length scales, see Figure \ref{su2Fit}. 

\begin{figure}[H]
	\centering
		\includegraphics[width=1\textwidth]{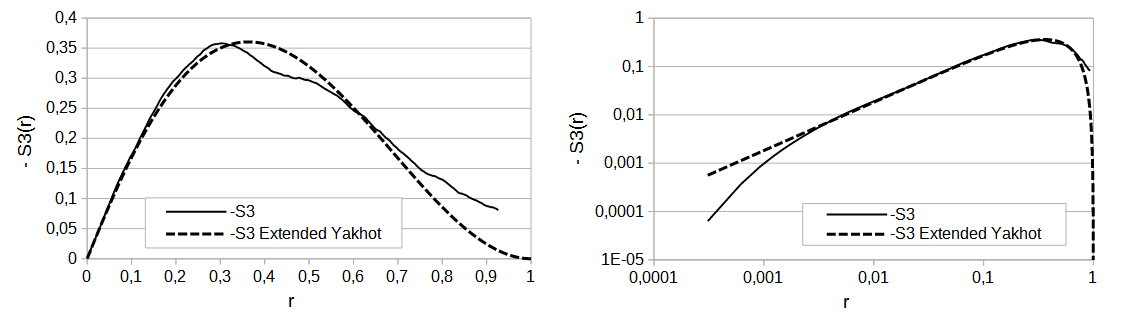}
	\caption{Third--order structure function as given by eq. (\ref{fullExtS3Solution}) for $\zeta_2=0.7$ (dashed lines) in comparison to experimental data (solid lines) in linear (left) and logarithmic (right) scale.}
	\label{su3Fit}
\end{figure}

Knowledge of $S_2(r)$ is the prerequisite to solving equation (\ref{extendedSunWithoutD}) for third order:
\begin{eqnarray}
	r \, \frac{\partial}{ \partial r} S_{3}(r) \;  & = &  \; \zeta_{3}\, S_{3}(r) \, + \, z_{3} \, r \, c(r) \, S_{2}(r). \label{extS3Eq} 
\end{eqnarray}
Inserting expression (\ref{extS2Solution}) for $S_2(r)$ we find the general solution
\begin{eqnarray}
	S_{3}(r) \;  & = &  \; K_3 \, r \, + \, 2 \, \frac{ z_3 }{1 - \zeta_2}  \, C \, r\, F(r),  \label{extS3Solution} 
\end{eqnarray}
where $K_3$ is the constant of integration and
\begin{eqnarray}
	F(r) \;  & = & \; \frac{1}{1+\zeta_2}  \, r^{1+\zeta_2} \, - \, \frac{\zeta_2}{2} \, r^2 \, - \, \frac{1}{2+\zeta_2} \, r^{2+\zeta_2} \, + \, \frac{\zeta_2}{3} \, r^3. \label{FvonR} 
\end{eqnarray}
$K_3$ is readily obtained from the four--fifth law (\ref{fourFifthWoDimensions}) as $K_3=-\frac{4}{5}\epsilon$ and the parameter $C$ can be determined from the condition that $S_3(r)$ is zero at the system scale:
\begin{eqnarray}
	C \;  & = &  
	  \; \frac{2}{5}\, \frac{\epsilon}{z_3} \, \frac{1-\zeta_2}{F(1)}. \label{valueOfC}
\end{eqnarray} 
The solution for the third order structure function finally simplifies to:
\begin{eqnarray}
	S_{3}(r) \;  & = &  \; - \, \frac{4}{5} \, \epsilon \, r \,  \left\{ \, 1 - \, \frac{F(r)}{F(1)} \right\}. \label{fullExtS3Solution} 
\end{eqnarray}
The prediction of the extended Yakhot model is found to be in good agreement with experimental data from the system scale through to the lower end of the inertial range, see figure \ref{su3Fit}. 

The parameter $C$ ultimately cancels out of the solution for $S_3(r)$ but enters the equations for higher order structure functions. The following approximation for this parameter can therefore be of interest: By expressing the second order scaling exponent $\zeta_2$ in terms of the small scaling anomaly $\delta$ defined as $\delta = \zeta_2 - 2/3$ we can expand the term $F(1)$ and obtain the following approximation of first order in $\delta$:
\begin{eqnarray}
	F(1) \; = \; \frac{1}{1+\zeta_2} \, - \, \frac{1}{2+\zeta_2} \, - \, \frac{\zeta_2}{6}  \; \approx \; \frac{1 \, - \, \zeta_2}{3} . \label{F1proxy}
\end{eqnarray}
This approximation is correct within a maximum relative deviation of $1.5\%$ in the relevant range\footnote{This range is defined by the commonly accepted experimental value for the scaling anomaly of $\delta = 0.029 \pm 0.004$ \cite{Arneodo}. This corresponds to a range of $0.692 \leq \zeta_2 \leq 0.700$ or $7 \leq A \leq 9$.} of parameters. With this approximation and the explicit expression (\ref{ZetaN}) for $z_3$ we obtain:
\begin{eqnarray}
	C \;  & = &  \; \frac{2}{5}\, \frac{\epsilon}{z_3} \, \frac{1-\zeta_2}{F(1)} \; = \; \frac{1}{5} \, A \, \epsilon \, \frac{1-\zeta_2}{F(1)} \nonumber \\[0.1cm]
		& \approx &  \frac{3}{5} \, A \, \epsilon  . \label{valueOfCproxy}
\end{eqnarray} 

\begin{figure}[H]
	\centering
		\includegraphics[width=1\textwidth]{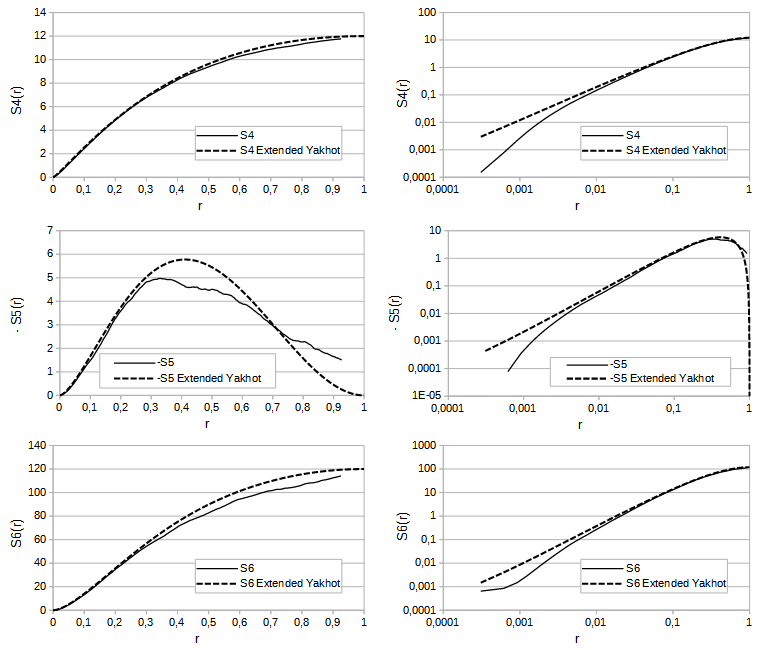}
	\caption{Structure functions (solid lines) of order four to six in comparison to the predictions of the extended Yakhot model (dashed lines) in linear (left) and logarithmic (right) scale.}
	\label{su4to6Fit}
\end{figure}

We conclude with a comparison of the (numerical) solutions of the model equation for structure functions of orders four to six with experimental data in figure \ref{su4to6Fit}. Considering that statistical noise increases with both length scale $r$ and order $n$ the agreement between model and experimental data can be considered good.

\section{Summary and Discussion}\label{discussion}
An extension of Yakhot's model of turbulence is proposed which introduces two additional parameters, $D$ and $C$, into the original equation for the probability density function $p(u,r)$ of the dimensionless velocity increment $u$:
\begin{equation}
	B \frac{\partial p}{\partial r} \, - \, \frac{\partial}{\partial u} 
	\left\{ u \frac{\partial p}{\partial r} \right\} \; = \; - \frac{A}{r} 
	\frac{\partial}{\partial u} \Big\{ u p \Big\} \, + \, 
	\frac{\partial^2}{\partial u^2} \Big\{ \, \big( \, u \, c(r) + D \, \big) \, p  \, \Big\}.
\end{equation}
where the function $c(r)$ is defined as
\begin{equation}
	c(r) \; = \; r \, (1-r) \, C \label{cDefinitionAgain}
\end{equation}
with $C$, as well as $A$, $B$ and $D$, being constant. 

Large scale boundary conditions and Kolmogorov's four--fifth allow to express the model parameters in terms of only one independent parameter. We find it most convenient to express $B$, $C$ and $D$ as functions of $A$:
\begin{eqnarray}	
	B \; & = & \; 3 \, (A-1) \nonumber \\
	C \; & \approx & \; \frac{3}{5} \, \epsilon \, A \nonumber \\
	D \; & = & \, - 2 \, A. \label{modelParameterization}
\end{eqnarray}
The exact expression (\ref{valueOfC}) for $C$ is slightly more involved but also depends on the parameter $A$ (and $\epsilon$) only. The model equation for the structure function $S_n(r)$ of order $n$ is obtained as
\begin{eqnarray}
	\frac{\partial}{ \partial r} S_{n}(r) \; & = & \; \frac{\zeta_{n}}{r} \, S_{n}(r) \, + \, z_{n} \, c(r) \, S_{n-1}(r) \, - \, 2(n-1) \,\zeta_n \, S_{n-2}(r), \nonumber \\[0.2cm]
	\zeta_{n} \;  & =  & \; \frac{A \, n}{3 \, (A-1) \, +n}, \nonumber \\
	z_{n} \;  & =  & \; \frac{n \, (n-1) }{3 \, (A-1) \, + \, n}. \label{extendedSunAgain}
\end{eqnarray}
For second and third order the solutions of eq. (\ref{extendedSunAgain}) are:
\begin{eqnarray}
	S_{2}(r) \;  & = &  \; \frac{2}{1-\zeta_2} \, \left( \, r^{\zeta_2} \, - \, \zeta_2 \, r \, \right), \label{ExtS2SolutionAgain} \\
	S_{3}(r) \;  & = &  \; - \, \frac{4}{5} \, \epsilon \, r \,  \left\{ \, 1 - \, \frac{F(r)}{F(1)} \right\}
\end{eqnarray}
with
\begin{eqnarray}
	F(r) \;  & = & \; \frac{1}{1+\zeta_2}  \, r^{1+\zeta_2} \, - \, \frac{\zeta_2}{2} \, r^2 \, - \, \frac{1}{2+\zeta_2} \, r^{2+\zeta_2} \, + \, \frac{\zeta_2}{3} \, r^3. 
\end{eqnarray}

The proposed extension of Yakhot's model clearly is of empirical nature. Yet, the additional terms can be motivated by physical arguments and in a straightforward manner be determined from very general considerations, namely the boundary conditions for the structure function's values and first order derivatives at the system length scale $L$. Remarkably, the extended model reproduces experimental data fairly well, from the system length scale through to the inertial range. In that context it is worth stressing (again) that none of the parameters has been used to optimize the fit to experimental data, including the free parameter $A$. This parameter is set to $7.3$ (corresponding to $\zeta_2=0.7$) following \cite{SreenivasanAndYakhot} where this value was identified as the limit of the scaling exponents $\zeta_n$ for $n \rightarrow \infty$ (cf. eq.(\ref{aAsZetaLimit})). 

When determined from large scale boundary conditions, the effective diffusion coefficient $D$ turns out to be proportional to $A$. It is this simple relation that makes the solution (\ref{ExtS2SolutionAgain}) for $S_2$ a straightforward extension of "conventional" inertial range scaling and establishes a link between the level of the structure function in the inertial range and the (known) large scale level. Having established this link is the main result of this paper.

Further investigations will have to follow. The most obvious indication for the proposed model not being the last word is the fact that both the function $c(r)$ as well as the assumption of a constant effective diffusion parameter $D$ violate the fundamental symmetry $p(-u,-r)=p(u,r)$ of the original model. This symmetry follows from properties of the Navier--Stokes equation \cite{Yakhot} and violating it hence constitutes a serious drawback. In order to restore this symmetry the diffusion coefficient $D$ has to be made a function of $r$ with $d(-r)=-d(r)$ and $c(r)$ needs to be adapted\footnote{These adaptions might require to give up the assumption of the extensions being elementary functions.} to fulfill $c(-r)=c(r)$. Conditions (\ref{cGoesToZero}) and (\ref{cAtOne}) on $c(r)$ are not per se in contradiction to this condition, but expression (\ref{valueOfD}) for $D$ will have to be rewritten as $d(r=+1) = - 2A$.

Despite the empirical nature of the approach and some remaining open questions, the results obtained with the proposed model for the transition from large to inertial range scales are unprecedented to the best of our knowledge. We therefore believe that the ideas presented here could serve as starting points for further theoretical and experimental investigations

\subsection*{Acknowledgments} 
We gratefully acknowledge fruitful discussions with J. Peinke and the provision of the high--quality dataset by courtesy of B. Castaing and B. Chabaud.

\appendix
\section{Experimental Data and Methods}\label{app}

The predictions of the original and the proposed extended Yakhot model are in this paper benchmarked against experimental data measured in a cryogenic axisymmetric helium gas jet at a Reynolds number of approx. $4 \cdot 10^5$. The data set contains $1.6 \cdot 10^7$ samples of the local velocity measured in the center of the jet. Taylor’s hypothesis of frozen turbulence is used to convert time lags into spatial displacements. Details on on the experimental setup and the data can be found in \cite{data, transition}.

Calculation of structure functions from the data is rather straightforward, but extracting reliable information about their large scale properties poses certain difficulties. In particular determining the system length scale $L$ is intricate owing to the facts that (i) large scale levels are approached with vanishing slopes and (ii) statistical noise can become considerable for large scales \footnote{For any given data series, the number of statistically independent observations of velocity increments decreases with increasing length scale\label{largeScaleDataIssue}}. The latter effect is distinctly more pronounced for odd--order structure functions where contributions with different signs sum up to zero. 

\begin{figure}[H]
	\centering
		\includegraphics[width=0.8\textwidth]{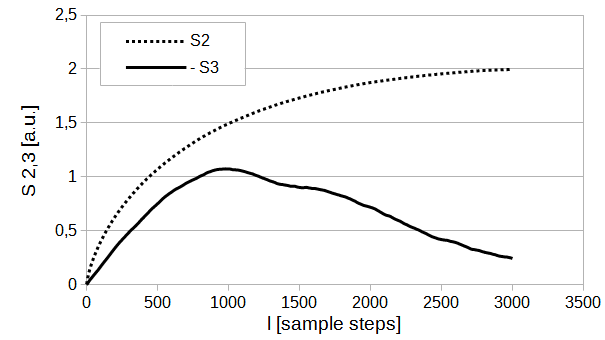}
	\caption{Empirical second (dotted line) and third (solid line) order structure function. The second order function has been normalized to a long--term level of $2,$ $-\mathscr{S}_3(l)$ has arbitrarily been scaled to a comparable order of magnitude for clarity of presentation. The length scale l is given in multiples of the sampling step size.}
	\label{sTwoAndThree}
\end{figure}

Both effects become apparent in figure \ref{sTwoAndThree} displaying the structure functions of order two and three calculated from the data set considered here. $\mathscr{S}_2$ clearly converges towards a constant level, but with the slope of the function decreasing for large scales the precise point of convergence cannot easily be identified. $\mathscr{S}_3$ shows the expected linear dependence on $l$ (see eq. (\ref{fourFifthsReduced})) for scales $l \ll L$, but even before the function reaches its maximum it exhibits clearly visible oscillations which are arguably a signature of statistical noise rather than a genuine physical effect.

\begin{figure}[H]
	\centering
		\includegraphics[width=0.8\textwidth]{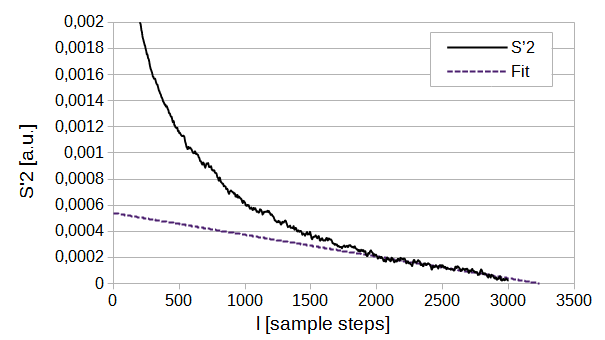}
	\caption{First--order derivative of the second order structure function (solid line) with linear fit (dashed line) in $2000 \leq l \leq3000$. The length scale l is given in multiples of the sampling step size.}
	\label{SystemLength}
\end{figure}

The method applied in this paper therefore builds upon the second order structure function. It furthermore makes use of the fact that the first order derivative of $S_2$ shows a more clearly pronounced convergence towards its large scale level (of zero) than the structure function itself, see figure \ref{SystemLength}, and hence determines $L$ as the scale for which the first order derivative of the second order structure function becomes zero.

For large scales the derivative of $S_2$ exhibits a range where it can be approximated by a linear function. For the data set considered here we obtain the system scale $L$ from linear fits as $L \approx 3250$ (sampling steps). The result varies slightly with the range used for the fit resulting in a relative error of $\approx 4\%$.

Two other length scales of interest are the lower and upper bounds of the inertial range. We define the inertial range as the range of length scales for which the four--fifth law (\ref{fourFifthsReduced}) is fulfilled. From eq. (\ref{fourFifthsReduced}) it follows that in the inertial range:
\begin{equation}
	\frac{l}{\mathscr{S}_3} \cdot \frac{\partial \mathscr{S}_3}{\partial l} \; = \; \zeta_3  \; = \; 1. \label{irDef}
\end{equation}

\begin{figure}[H]
	\centering
		\includegraphics[width=0.8\textwidth]{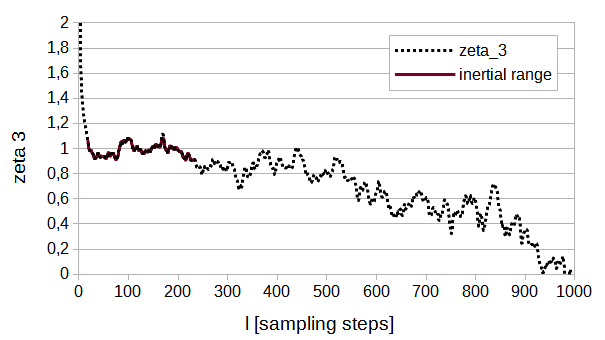}
	\caption{$\zeta_3$ as defined by eq. (\ref{irDef}). The solid line indicates the inertial range where experimental values for $\zeta_3$ deviate by less than $10\%$ from the theortical value of $1$.}
	\label{inertialRangeDef}
\end{figure}

Figure \ref{inertialRangeDef} shows the left--hand side of eq. (\ref{irDef}) determined from experimental data. The bold curve indicates the range of values that deviate by less than $10\%$ from the theoretical level of $1$. This criterion is complemented by the additional requirement that the average of values in this range must not deviate from the expected level of $1$ by more than the standard deviation of values. This is the case with the average value and the standard deviation of $\zeta_3$ in the marked range being $0.98$ and $0.05$, respectively. We obtain the lower and upper bounds of the inertial range as $l=20$ and $l=230$.

\end{document}